\documentclass[pdflatex,sn-apa]{sn-jnl}

\usepackage{graphicx}%
\usepackage{multirow}%
\usepackage{amsmath,amssymb,amsfonts}%
\usepackage{amsthm}%
\usepackage{mathrsfs}%
\usepackage[title]{appendix}%
\usepackage{xcolor}%
\usepackage{textcomp}%
\usepackage{manyfoot}%
\usepackage{booktabs}%
\usepackage{algorithm}%
\usepackage{algorithmicx}%
\usepackage{algpseudocode}%
\usepackage{listings}%
\usepackage{natbib}
\usepackage[normalem]{ulem} 

\usepackage{subcaption}
\usepackage{wrapfig}
\usepackage{amsmath}
\usepackage{lipsum} 
\usepackage{multicol} 

\begin{document}

\title[One year of ASPEX-STEPS observations]{One Year of ASPEX-STEPS Operation: Characteristic Features, Observations and Science Potential}

\author*[1]{\fnm{Jacob} \sur{Sebastian}}\email{jacobs@prl.res.in}
\author[1]{\fnm{Bijoy} \sur{Dalal}}
\author[1,2]{\fnm{Aakash} \sur{Gupta}}
\author[1]{\fnm{Shiv Kumar} \sur{Goyal}}
\author[1]{\fnm{Dibyendu} \sur{Chakrabarty}}
\author[1]{\fnm{Santosh V.} \sur{Vadawale}}
\author[1]{\fnm{M.} \sur{Shanmugam}}

\author[1]{\fnm{Neeraj Kumar} \sur{Tiwari}}

\author[1]{\fnm{Arpit R.} \sur{Patel}}
\author[1]{\fnm{Aveek} \sur{Sarkar}}
\author[1]{\fnm{Aaditya} \sur{Sarda}}
\author[1]{\fnm{Tinkal} \sur{Ladiya}}
\author[1]{\fnm{Prashant} \sur{Kumar}}
\author[1]{\fnm{Manan S.} \sur{Shah}}
\author[1]{\fnm{Abhishek} \sur{Kumar}}
\author[1,2]{\fnm{Shivam} \sur{Parashar}}
\author[1]{\fnm{Pranav R.} \sur{Adhyaru}}
\author[1]{\fnm{Hiteshkumar L.} \sur{Adalja}}
\author[1]{\fnm{Piyush} \sur{Sharma}}
\author[1]{\fnm{Abhishek J.} \sur{Verma}}
\author[1]{\fnm{Nishant} \sur{Singh}}
\author[1]{\fnm{Sushil} \sur{Kumar}}
\author[1]{\fnm{Deepak Kumar} \sur{Painkra}}
\author[1]{\fnm{Swaroop B.} \sur{Banerjee}}
\author[1]{\fnm{K. P.} \sur{Subramaniam}}
\author[3]{\fnm{Bhas} \sur{Bapat}}
\author[1]{\fnm{M. B.} \sur{Dadhania}}
\author[1]{\fnm{P.} \sur{Janardhan}}
\author[1]{\fnm{Anil} \sur{Bhardwaj}}

\affil*[1]{\orgname{Physical Research Laboratory}, \city{Ahmedabad}, \postcode{380009}, \country{India}}

\affil[2]{\orgname{Indian Institute of Technology Gandhinagar}, \city{Gandhinagar}, \postcode{382055}, \country{India}}

\affil[3]{\orgname{Indian Institute of Science Education and Research}, \city{Pune}, \postcode{411008}, \country{India}}


\abstract{The SupraThermal and Energetic Particle Spectrometer (STEPS), a subsystem of the Aditya Solar wind Particle EXperiment (ASPEX) onboard India’s Aditya-L1 satellite, is designed to study different aspects of energetic particles in the interplanetary medium from the Sun-Earth L1 point using six detector units oriented in different directions. This article presents details of the one-year operation (08 January 2024 - 28 February 2025) of the AL1-ASPEX-STEPS after the insertion of the satellite into the final halo orbit around the L1 point with emphasis on performance, science observations, and scientific potentials. Four out of six AL1-ASPEX-STEPS units exhibit a stable detector response throughout the observation period, confirming operational robustness. This work also includes the temporal variation of particle fluxes, spectra of ions during selected quiet times and transient events, and cross-comparisons with existing instruments at the L1 point. A strong correlation (with coefficient of determination, R$^2 \sim 0.9$) is observed in the cross-comparison study, establishing the reliability of the AL1-ASPEX-STEPS observations. AL1-ASPEX-STEPS also captures different forms of energetic ion spectra similar to those observed by previous missions. These results underscore the instrument’s potential to contribute significantly to the study of energetic particle acceleration, transport, and long-term space weather monitoring from the Sun–Earth L1 vantage point.}


\maketitle

\section{Introduction}
 Given the current advancements in space exploration, it is necessary to assess the space environment beforehand. One of the major harmful agents to the electronics \citep{iucci2005space} on-board space assets and humans in space is Solar Energetic Particles (SEPs) having energies ranging from a few hundreds of keVs to a few hundreds of MeVs. These highly energetic particles (or ions) are mostly generated from the Sun or within the heliosphere. While impulsive SEPs are thought to be generated at the reconnection sites corresponding to solar flares \citep{kahler2001coronal}, gradual SEPs are mostly generated by the interplanetary (IP) shocks associated with coronal mass ejections (CMEs). Stream/corotating interaction regions (SIR/CIRs) also are observed to be associated with energetic particles at various heliocentric distances from the Sun \citep{richardson1984low}. As per present understanding, SEPs are accelerated from a suprathermal (with energies from 10s of keV to few MeV) ion pool in the IP medium \citep{gloeckler2003ubiquitous}. However, the nature, sources, and generation mechanisms of these ``seed'' particles are not fully comprehensive.    
 
Several space missions have been dedicated to measuring suprathermal and energetic particles. The Wind spacecraft carries the Suprathermal Energetic Particle (STEP) instrument, part of the Energetic Particles: Acceleration, Composition, and Transport (EPACT) suite \citep{von1995energetic}. Likewise, the ACE spacecraft is equipped with the Electron, Proton, and Alpha Monitor (EPAM, \citealp{gold1998electron}), the Solar Isotope Spectrometer (SIS, \citealp{stone1998solar}), and the Ultra-Low Energy Isotope Spectrometer (ULEIS, \citealp{mason1998ultra}), all of which measure particles in the suprathermal and energetic particle domain. While these missions have provided valuable insights, their spin-stabilized platforms limit the ability to retain directional information unless the data is processed in a specific manner. The twin Solar TErrestrial RElations Observatory-Ahead and Behind (STEREO-A and B) missions, on the other hand, are three-axis stabilized and move in different heliocentric orbits (STEREO-B stopped observations from 2014). The Solar Energetic Particle (SEP, \citealp{mason2008suprathermal}) suite of the In situ Measurements of Particles And CME Transients (IMPACT, \citealp{luhmann2005impact}) investigation on STEREO is capable of measuring suprathermal and SEPs for a wide range of energies and species. However, these observations are not from the Sun-Earth first Lagrange (L1) point. 

In contrast, the Suprathermal and Energetic Particle Spectrometer (STEPS, \citealp{goyal2018aditya, goyal2025aditya}), a sub-system of the Aditya Solar wind Particle EXperiment (ASPEX) onboard India’s first dedicated solar mission, Aditya-L1 (AL1, \citep{seetha2017aditya, tripathi2022aditya}), offers a significant advantage. Mounted on a three-axis stabilized spacecraft, AL1-ASPEX-STEPS enables direction-resolved measurements across a broader energy range, which could enhance our understanding of particle acceleration and transport in the inner heliosphere. The primary science goals of AL1-ASPEX-STEPS include but not limited to the investigations on (i) the origin of suprathermal particles in the IP medium, (ii) SEP characterization, (iii) connection between suprathermal particles and SEPs, (iv) SIR/CIR-accelerated energetic particles, and (v) energetic ions upstream of the Earth's bow shock. 

AL1 is currently operating in a halo orbit around the L1 point, enabling continuous observation of the Sun. Launched on 02 September 2023, the mission performed four Earth-bound orbit-raising maneuvers, followed by a trans-L1 cruise phase, and was successfully inserted into its final orbit on 06 January 2024. AL1-ASPEX-STEPS was the first scientific payload to become operational on 10 September 2023 and has been continuously recording particle data throughout the mission, starting from its passage through the Earth’s magnetosphere, during the trans-L1 cruise phase, and now at the halo orbit around the L1 point. The initial phase of AL1-ASPEX-STEPS operations focused on validating the instrument’s performance by comparing in-flight measurements with pre-flight calibration data. This included evaluating housekeeping parameters, calibration pulses, and raw trigger counts data to ensure that the payload functions as expected. During the performance verification (PV) phase, configuration changes were implemented to optimize the performance of AL1-ASPEX-STEPS. 

This paper presents the one-year performance summary of AL1-ASPEX-STEPS following the halo orbit insertion of Aditya-L1. Section 2 provides a brief description of the instrument and its datasets. Section 3 details the performance of AL1-ASPEX-STEPS during the observation period. Section 4 discusses the scientific observations and potential of the instrument, while Section 5 summarizes its overall science prospects.

\section{Instrumentation and datasets}
\begin{figure}[ht!]
\centering
\includegraphics[width=0.5\linewidth]{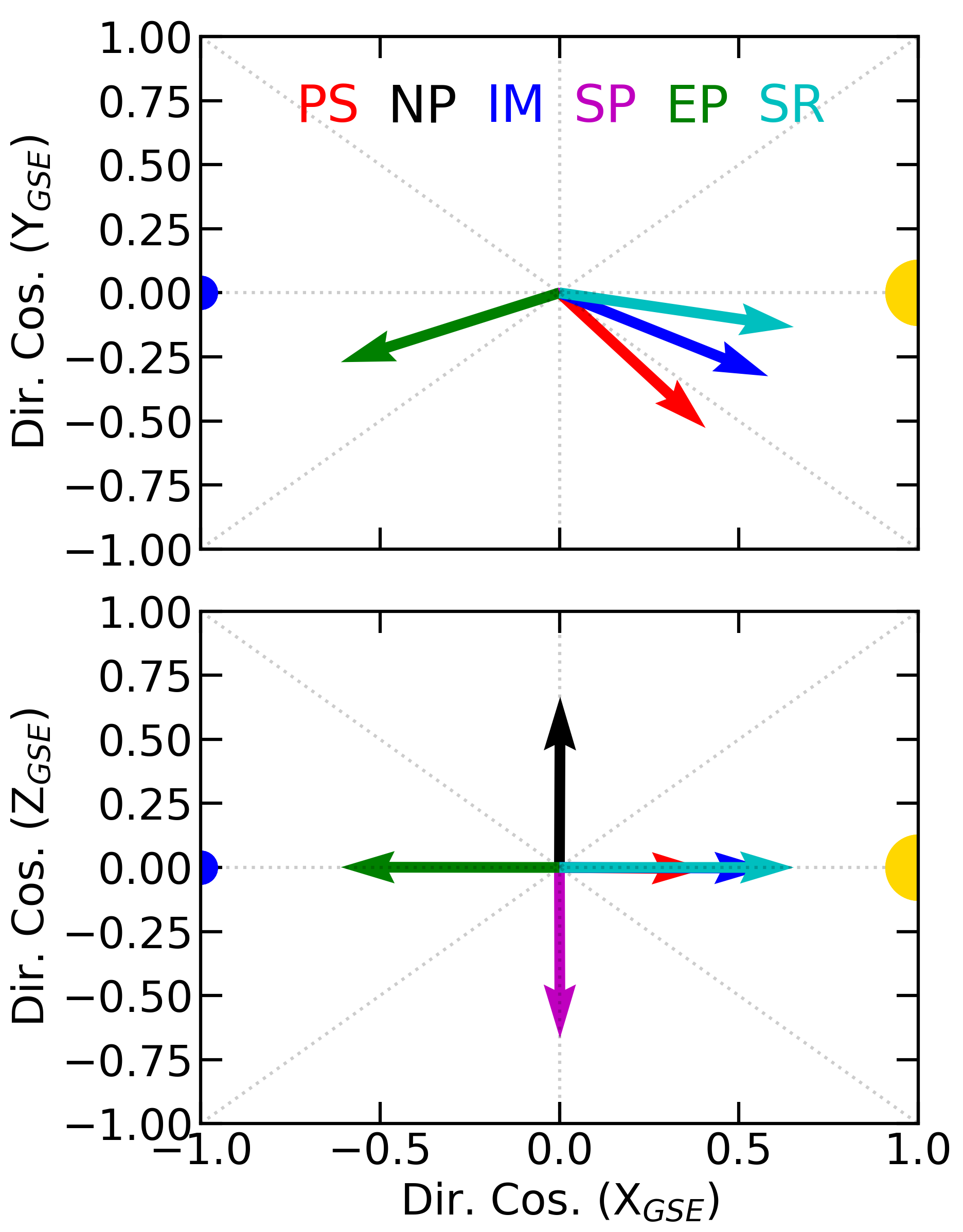}
\caption{Nominal orientations of PS (red), NP (black), IM (blue), SP (magenta), EP (green), and SR (cyan) detector units of AL1-ASPEX-STEPS with respect to the Geocentric Solar Ecliptic (GSE) coordinate system. The lengths of the arrows are determined based on the direction cosines plotted in the respective panels. It can be seen that SR, IM, PS, ad EP units are in the ecliptic plane. On the other hand, NP and SP units are oriented along the ecliptic north and south directions, respectively. The Sun and the Earth are represented symbolically by the yellow and blue filled circles, respectively. \label{fig_1_1_BD}}
\end{figure}

AL1-ASPEX-STEPS \citep{goyal2025aditya} consists of six detector units, each oriented in a distinct direction to sample particles (ions) from a wide range of incident angles. These units are designated as Sun Radial (SR), Parker Spiral (PS), InterMediate to SR and PS (IM), Earth Pointed (EP), North Pointed (NP), and South Pointed (SP). The nominal configurations of AL1-ASPEX-STEPS detector units are shown in Figure \ref{fig_1_1_BD}, where we plot direction cosines of angles made by the boresight (axis of a cone) of PS, NP, IM, SP, EP, and SR detector units with the axes of the Geocentric Solar Ecliptic (GSE) coordinate system. As can be seen in Figure \ref{fig_1_1_BD}, SR, IM, PS, and EP units are in the ecliptic plane. On the other hand, NP and SP units are looking at the ecliptic north and south directions, respectively. This configuration is generally maintained for most of the mission life unless there is a specific operational maneuver of the spacecraft. 

Each detector unit of AL1-ASPEX-STEPS uses fully depleted Si-PIN detector generating electron-hole pairs, proportional to the energy of the incident particles. The SR, PS, and EP units employ detectors with a thickness of $300\,\mu\mathrm{m}$ and feature a dual-window configuration. This configuration includes two active regions with different dead layer thicknesses: approximately $0.1\,\mu\mathrm{m}$ for the inner region (7 mm diameter) and $0.8\,\mu\mathrm{m}$ for the outer region (7-18 mm diameter). In contrast, IM, NP, and SP units utilize a single-window configuration with a detector thickness of $250\,\mu\mathrm{m}$ with 0.2 $\mu$m dead layer. The variation in dead layer thickness affects the energy deposited by incoming particles in the silicon detector. To capture a wide range of particle energies, each detector is equipped with two parallel signal processing chains: high-gain (HG) and low-gain (LG). Particle energies are measured in 256 analog-to-digital converter (ADC) channels among which channels 0–127 are assigned to the HG chain (hereafter, HG channels) covering up to a typical energy of 2 MeV and Channels 128–255 assigned to the LG chain (hereafter, LG channels) for higher energies. More instrument details are available in  \cite{goyal2018aditya} and \cite{goyal2025aditya}. AL1-ASPEX-STEPS data, after the completion of the performance verification phase, are available at \url{https://pradan.issdc.gov.in/al1/}.  

\section{Instrument performance}
AL1-ASPEX-STEPS has been operational since 10 September 2023. The performance (e.g. stability of instrument health parameters) of the detectors has been continuously monitored by the payload team. Some of the performance aspects are covered in the following sub-sections.  

\subsection{Performance of calibration pulses on-board}
The performance of the HG channels of AL1-ASPEX-STEPS detectors is periodically assessed on-board using calibration pulses integrated into SR, PS, and SP units. 

\begin{figure}[ht!]
\centering
\includegraphics[width=0.8\linewidth]{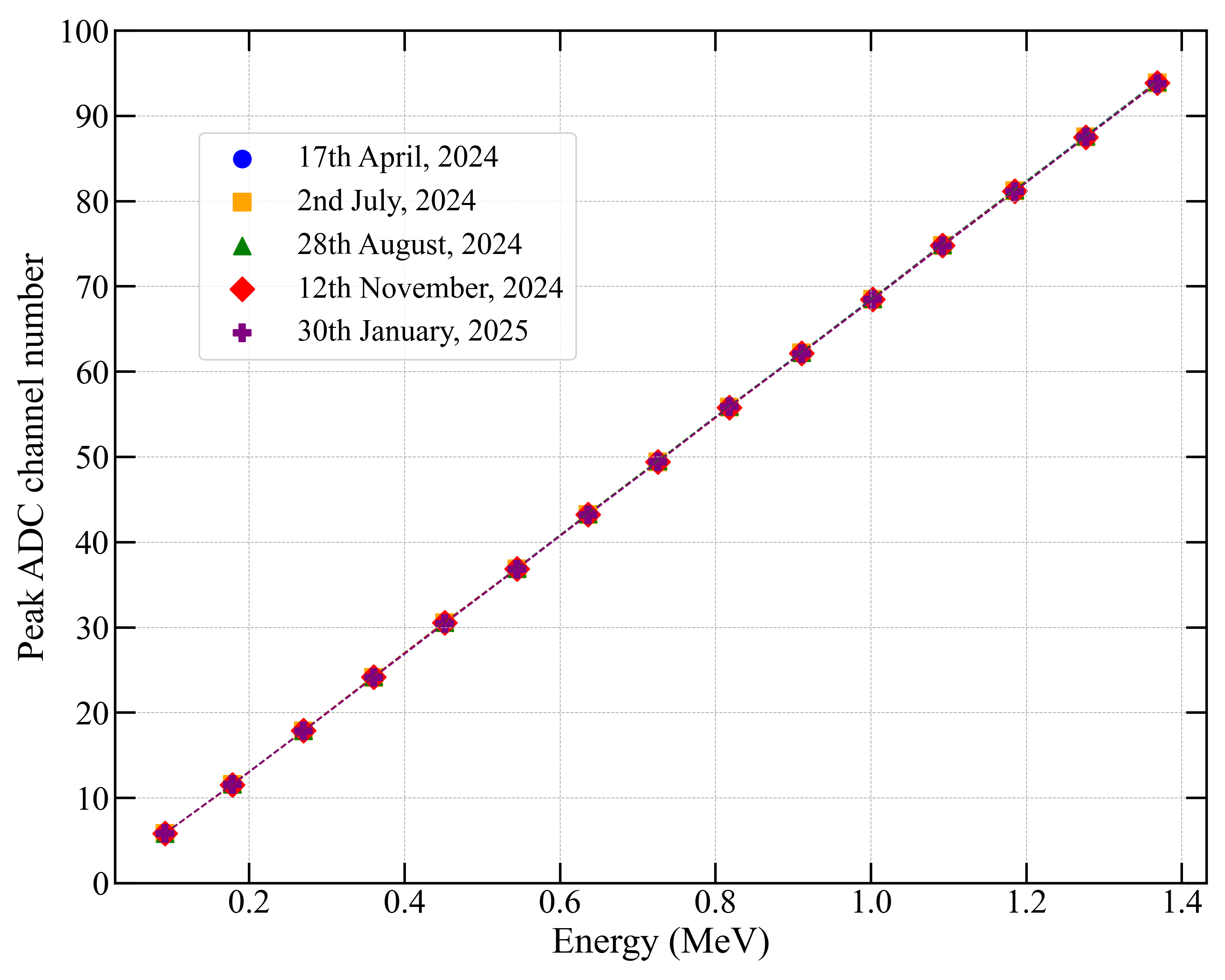}
\caption{Comparison of calibration measurements recorded at multiple epochs for the PS detector unit. Note the consistency of the electronics gain for different epochs. \label{fig_1_JS}}
\end{figure}

In this method, a known amount of charge is given to the input of detector electronics in terms of calibration pulses, simulating the electron-hole pair generation. The resulting spectra are analyzed to monitor any variations in the electronics gain over time. Calibration pulses acquired during the pre-launch and immediately after the halo orbit insertion have already been studied \citep{goyal2025aditya}. In this work, we extend the analysis to calibration pulses recorded from that period up to February 2025.

Figure \ref{fig_1_JS} presents the comparison for PS unit, where the X-axis denotes the deposited energy corresponding to the calibration pulses and the Y-axis shows the ADC channels corresponding to the centroids (of the Gaussian fits) of the calibration pulses. The maximum deviation in the ADC channels is around 0.2. This minor variation is likely due to the different temperatures encountered by the instrument during various measurements. Importantly, the slope of the calibration curve remains unchanged across different data points, indicating that the electronics performance is stable. Consequently, no update to the calibration database is required at this stage.


Regular monitoring will continue and updates to the calibration database will be made if future analyses indicate significant deviations. Calibration pulse measurements of SR and SP detector units are unavailable due to the saturation of detectors. This aspect is discussed in detail later in this work.

\subsection{Performance of Health monitoring parameters over time}
\begin{figure}[ht!]
\centering
\includegraphics[width=0.9\linewidth]{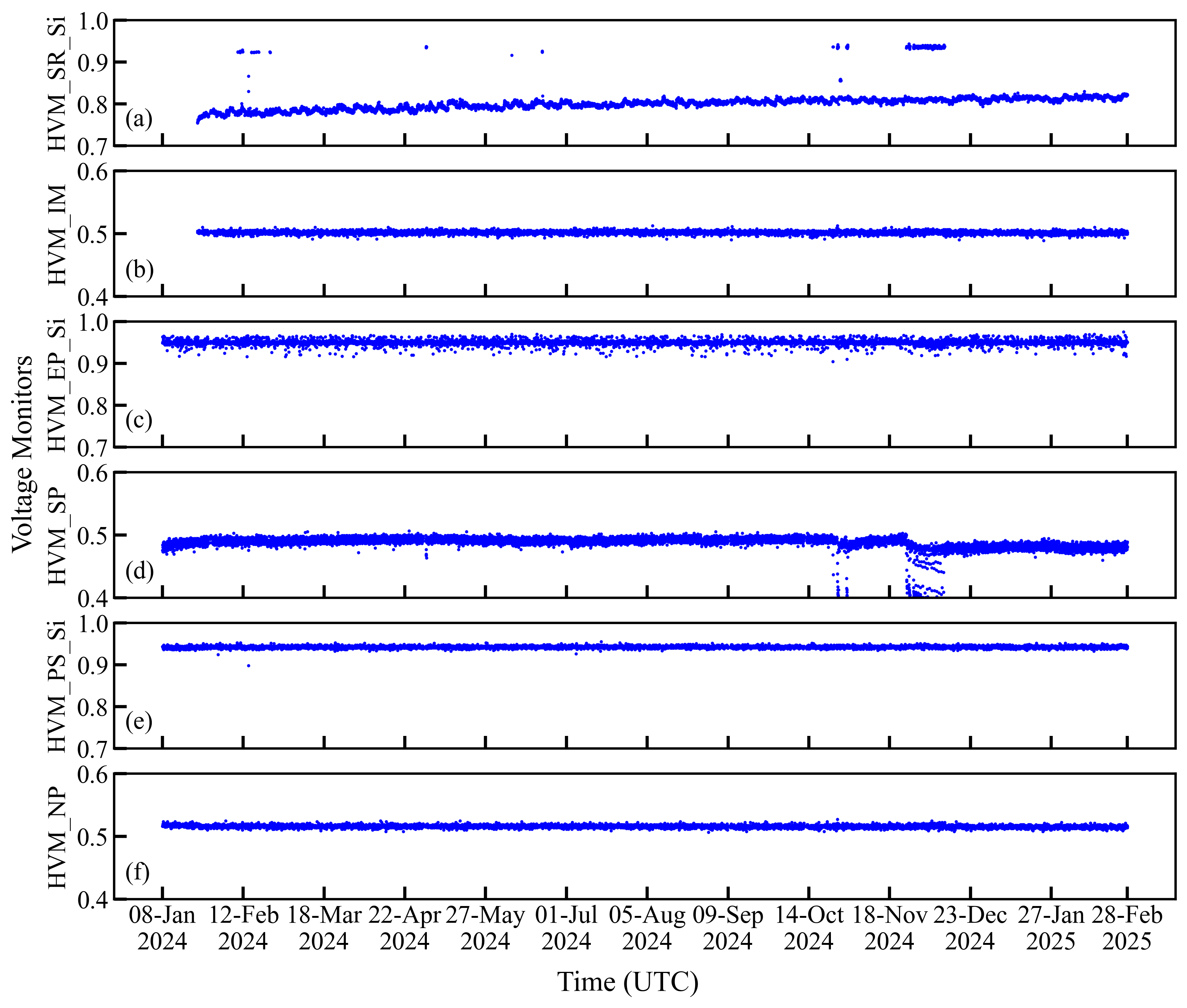}
\caption{Time series plot of housekeeping parameters, primarily the high-voltage monitor (HVM) readings of the detector units, from 8 January 2024 to 28 February 2025 \label{fig_2_JS}}
\end{figure}

The health of AL1-ASPEX-STEPS units can be assessed through the continuous monitoring of the detector voltages called the High Voltage Monitors (HVM). Under nominal operating conditions, expected HVM values are $\approx$ 0.95 V for the SR, PS, and EP units, and $\approx$ 0.52 V for the IM, NP, and SP units. These nominal values are obtained during the functionality test performed before the launch. A drop in HVM values indicates current loading possibly due to saturation effects, which in turn may lead to performance degradation.

Figure \ref{fig_2_JS} shows HVM values for the Si-PIN detectors of SR, IM, EP, SP, PS, and NP units from 08 January 2024 to 28 February 2025. IM, EP, PS, and NP units consistently maintain nominal HVM values throughout this period. In contrast, Si-PIN detectors of SR and SP units exhibit drops in their respective HVM values. The SP unit typically shows a reduced HVM value around 0.48 V, while the SR unit demonstrates a more substantial drop to 0.83 V. These persistent reductions indicate saturation effects and associated current loading in the SR and SP units. In addition, the occasional abrupt changes in the HVM values correspond to spacecraft rotation events, which briefly alter the incident ion flux and thus affect detector loading. These points are explained in the next section. 

\subsection{Saturation of SR and SP units of AL1-ASPEX-STEPS}
Out of the six units in AL1-ASPEX-STEPS, four -- PS, EP, IM, and NP -- are providing good-quality data which are being used for science. However, SR and SP units are affected by saturation, so their data cannot currently be used for scientific analysis.

\subsubsection{Saturation of SR unit}
In the nominal configuration, the spacecraft’s +yaw axis is aligned toward the Sun, meaning the +yaw-to-Sun angle is 0°. The SR unit was designed to measure particles arriving from the solar radial direction. To block direct sunlight from entering the detector, the unit was equipped with collimators and mounted with a small offset of $1.5^\circ$ from the +yaw axis. 

\begin{figure}[ht!]
\centering
\includegraphics[width=0.8\linewidth]{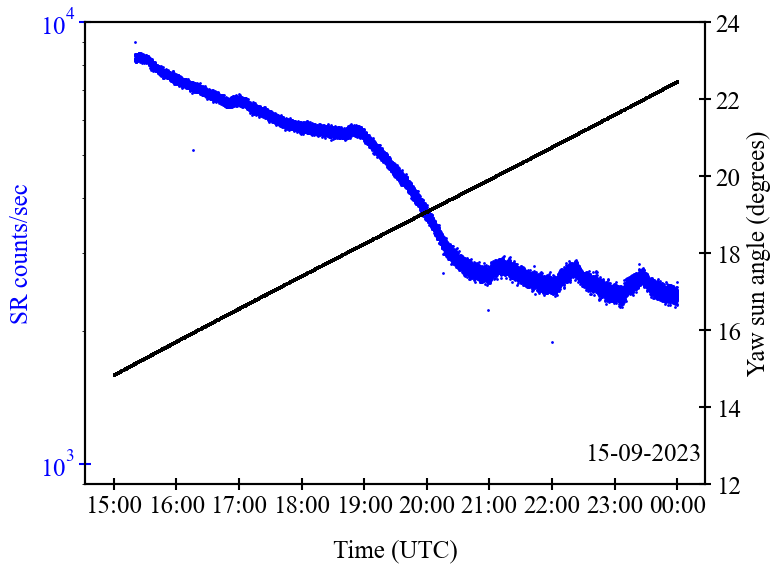}
\caption{Time series plots showing SR count rates (counts/sec) and the +Yaw angle to the Sun. A decrease in count rates is observed with increasing angle \label{fig_3_JS}}
\end{figure}

However, despite these measures, the detector response shows signs of saturation due to scattered sunlight leaking into the system. This effect is evident in the HVM signal from the SR unit (Figure \ref{fig_3_JS}, panel a), where the voltage, ideally expected to be around 0.95 V, drops to approximately 0.83 V as the detector becomes loaded. This issue was identified during the sun-simulator test, but a conscious decision was made to keep the detector’s aperture open to allow scanning of the magnetosphere during the Earthbound phase of the mission.

SR detector performs nominally when the spacecraft rotates such that sunlight no longer reaches the detector. As shown in Figure \ref{fig_3_JS}, when the +yaw-to-Sun angle increases, i.e., SR looks gradually away from the Sun, a corresponding decrease in SR counts is observed. From Figure \ref{fig_3_JS}, it can be seen that when +yaw-to-Sun angle $<\approx$ $16.5^\circ$ the detector gets saturated. The sudden spikes to 0.95 V observed in the HVM SR plots, particularly around November–December 2024, correspond to spacecraft rotations during which scattered sunlight is no longer incident on the detector, thus preventing current loading and allowing the detector to recover its nominal voltage.

Since SR and IM units share common front end electronics, switching off SR unit would automatically switch off IM unit. In order to maximize the output from AL1-ASPEX-STEPS, SR unit has been kept operational with a higher low-level discriminator (LLD) cut-off at the instrument level. Therefore, data from SR unit can not be used for scientific analyses in its nominal orientation at L1.

\subsubsection{Saturation of SP detector unit}
\begin{figure}[ht!]
\centering
\includegraphics[width=0.9\linewidth]{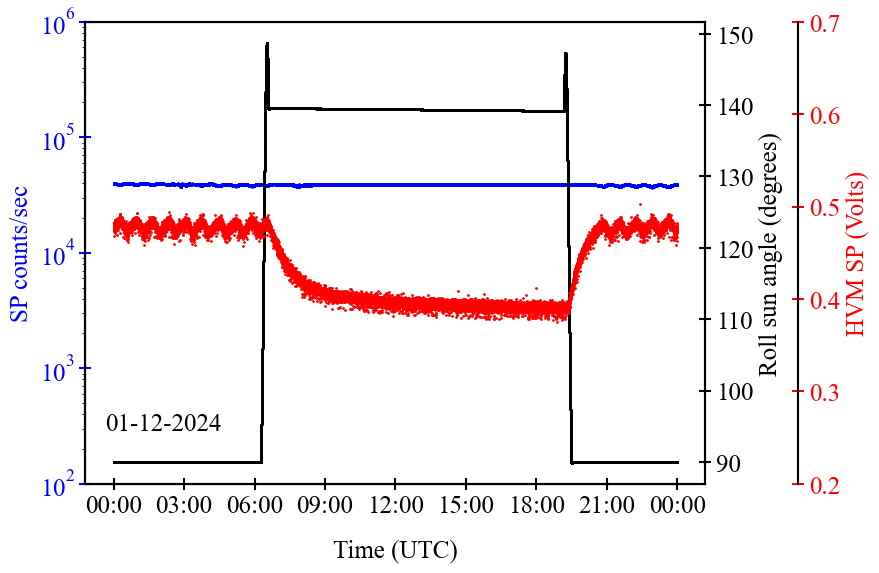}
\caption{Time series plots showing SP count rates (counts/sec) alongside the +Roll angle to the Sun and the high-voltage monitor (HVM) readings of the SP detector unit. Variations in spacecraft orientation do not affect the observed saturation. \label{fig_4_JS}}
\end{figure}

SP detector unit is oriented along the spacecraft’s –roll axis, and under nominal conditions, it is not expected to receive direct illumination from any light source. However, during the initial payload operations, the count rate measured by SP unit was getting saturated although HVM indicated only a small but consistent load -- shifting from 0.52 V to 0.48 V -- (see panel (d) of Figure \ref{fig_2_JS}). This aspect is further illustrated in Figure \ref{fig_4_JS} where we show variations of count rates (blue), +roll-to-Sun angle of the spacecraft (black), and HVM values (red) of SP unit on 01 December 2024. As can be seen in Figure \ref{fig_4_JS}, AL1 was in a rotated configuration from 06:00 UTC to 19:30 UTC, during which the spacecraft's +roll-to-Sun angle changed from $90^\circ$ to $140^\circ$.    

Throughout this operation, the detector count rate remained unchanged, while HVM value showed a decrease from 0.48 V to 0.4 V, indicating increased current loading and persistent saturation. Upon returning to the original spacecraft configuration after 19:30 UTC, HVM value reverted to 0.48. There was no corresponding change in the detector count rates. This behavior confirms that the saturation in SP unit is largely unaffected by the spacecraft's orientation. A likely cause for this persistent saturation could be the primary reflection of sunlight from the engine cone along with multiple secondary reflections from the multi-layer insulation (MLI) surrounding the detector unit.

Since the EP and SP units share common front-end electronics, switching off the SP unit would switch-off the EP unit as well. To ensure continued EP operation, the SP unit is kept on with a higher LLD cut-off, rendering its data unsuitable for scientific analysis due to saturation effects. Therefore, only data from the PS, EP, IM, and NP units— which are functioning as expected— are presented in this work. 

\subsection{Incorporation of gain toggling mode}
As mentioned in Section 2, AL1-ASPEX-STEPS electronics system uses an analog multiplexer with two parallel gain chains -- HG and LG. In the ``default'' mode, the analog multiplexer dynamically selects between HG and LG based on the incident signal amplitude. However, this mode introduced a discontinuity: a range of ADC channels near the transition between HG and LG (starting from ADC channel No. 128) often remains unpopulated due to timing uncertainties in multiplexer switching. This limited the usable energy range of ion spectra up to $\approx$ 1.3 MeV for IM and NP and 2 MeV for PS and EP units. 

To overcome this limitation, a ``toggling'' mode was tested during PV phase. In this mode, the analog multiplexer alternates between HG and LG every 5 minutes, allowing uninterrupted data collection in each gain setting. This approach eliminates the energy gap and enables reconstruction of a continuous energy spectrum extending beyond 1.3 MeV for IM and NP and beyond 2 MeV for PS and EP. Following successful validation, toggling mode was adopted for continuous operation starting from 11 May 2024 onwards. Although this mode reduces the temporal resolution from 1 second to 10 minutes, it significantly enhances the usable energy range. This temporal resolution is reasonable for energetic particles at L1 point. It is to be noted that within the 5 minutes of data (in either HG or LG), data is provided for every 1 second.

\section{Observations}
As stated above, PV phase extended till 10 May 2024 followed by Regular Science Observations (RSO) phase in ``toggling'' mode starting from 11 May 2024. In this section, we present different aspects of AL1-ASPEX-STEPS observations during 08 January 2024 -- 28 February 2025. This time period includes both PV phase and RSO phase observations.           
\begin{figure}[ht!]
    \centering
    \begin{subfigure}[t]{1.0\textwidth}
        \centering
        \includegraphics[width=\linewidth]{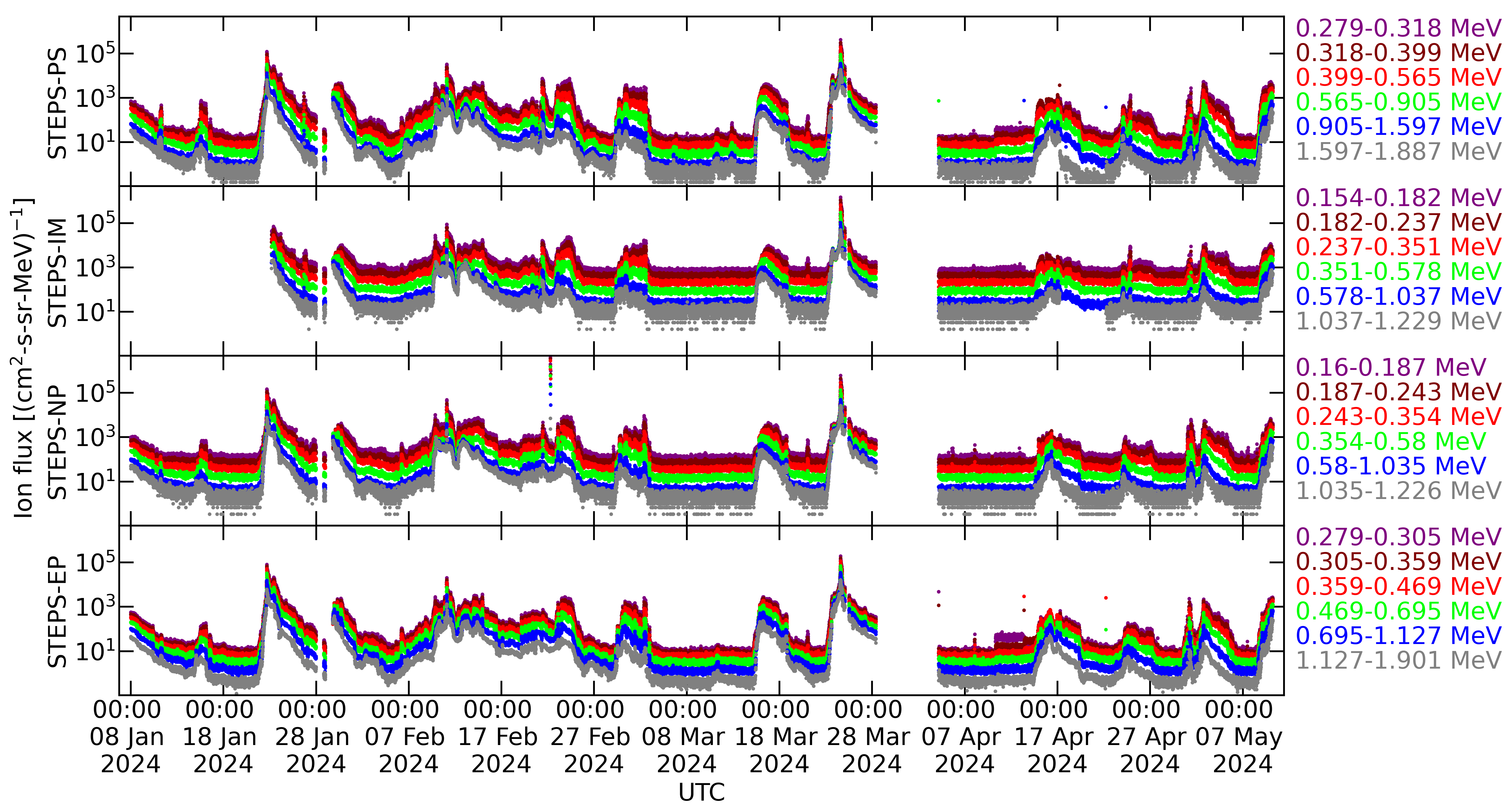} 
        \caption{}
        \label{fig_1a_BD}
    \end{subfigure}\hfill
    \begin{subfigure}[b]{1.0\textwidth}
        \centering
        \includegraphics[width=\linewidth]{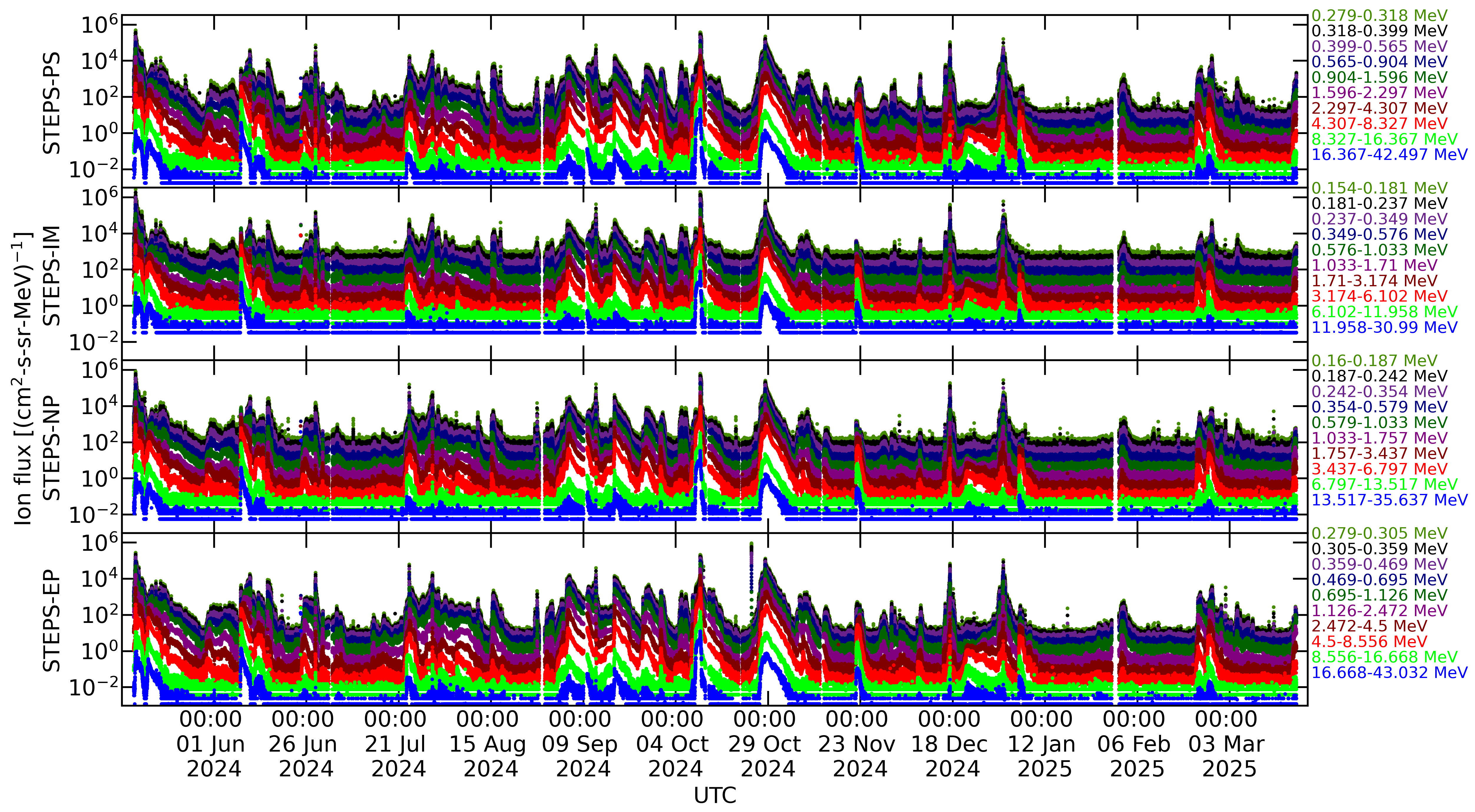} 
        \caption{}
        \label{fig_1b_BD}
    \end{subfigure}
\caption{Light curves of ion fluxes at different energy bands (a) before toggling (during 08 January 2024 00:00 UT - 09 May 2024 23:59 UT) and (b) after toggling (during 10 May 2024 00:00 UT - 21 March 2025 23:59 UT). Note that the energy bins are created based on the HG ADC channels in (a) and both the HG and LG ADC channels in (b). \label{fig_1_BD}}
\end{figure}
\subsection{Temporal variations of ion fluxes}
Figure \ref{fig_1a_BD} shows the temporal variations of 10 minutes averaged ion fluxes at different energy bands as observed by PS, IM, NP, and EP detector units (from top to bottom) of AL1-ASPEX-STEPS during 08 January 2024 00:00 UTC - 09 May 2024 23:59 UTC (i.e. part of PV phase). Although PS and EP have inner and outer detectors, ion fluxes from the outer detectors of these two units are shown here. This is because the inner detectors display distinct spectral signatures, and the deconvolution of their features is still in progress.

As stated in Section 3.4, AL1-ASPEX-STEPS units were operated in the ``default'' mode during this interval. The reported ion fluxes in Figure \ref{fig_1a_BD} are therefore obtained from HG channels (ADC channel No. 0 - 127) of the AL1-ASPEX-STEPS sensors and the energies are restricted to $\approx$ 1.3 MeV for IM and NP and to $\approx$ 2 MeV for PS and EP. The ion fluxes at different energy channels constituted from the HG and LG ADC channels after switching on the ``toggling'' mode are shown in Figure \ref{fig_1b_BD}. As can be seen, the variations in the ion fluxes for higher energies are captured. 

Figures \ref{fig_1a_BD} and \ref{fig_1b_BD} show multiple transient enhancements (events) in the energetic ion fluxes above a quiet background. These enhancements are generally associated with transient events like interplanetary (IP) shocks driven by interplanetary coronal mass ejections (ICMEs) and stream/corotating interaction regions (SIR/CIRs). However, we do not characterize these flux-enhancement events in this work. Instead, we discuss various features observed by different AL1-ASPEX-STEPS units. For this purpose, we focus on ion fluxes observed by different AL1-ASPEX-STEPS units before and during an event. 

\begin{figure}[ht!]
\centering
\includegraphics[width=\linewidth]{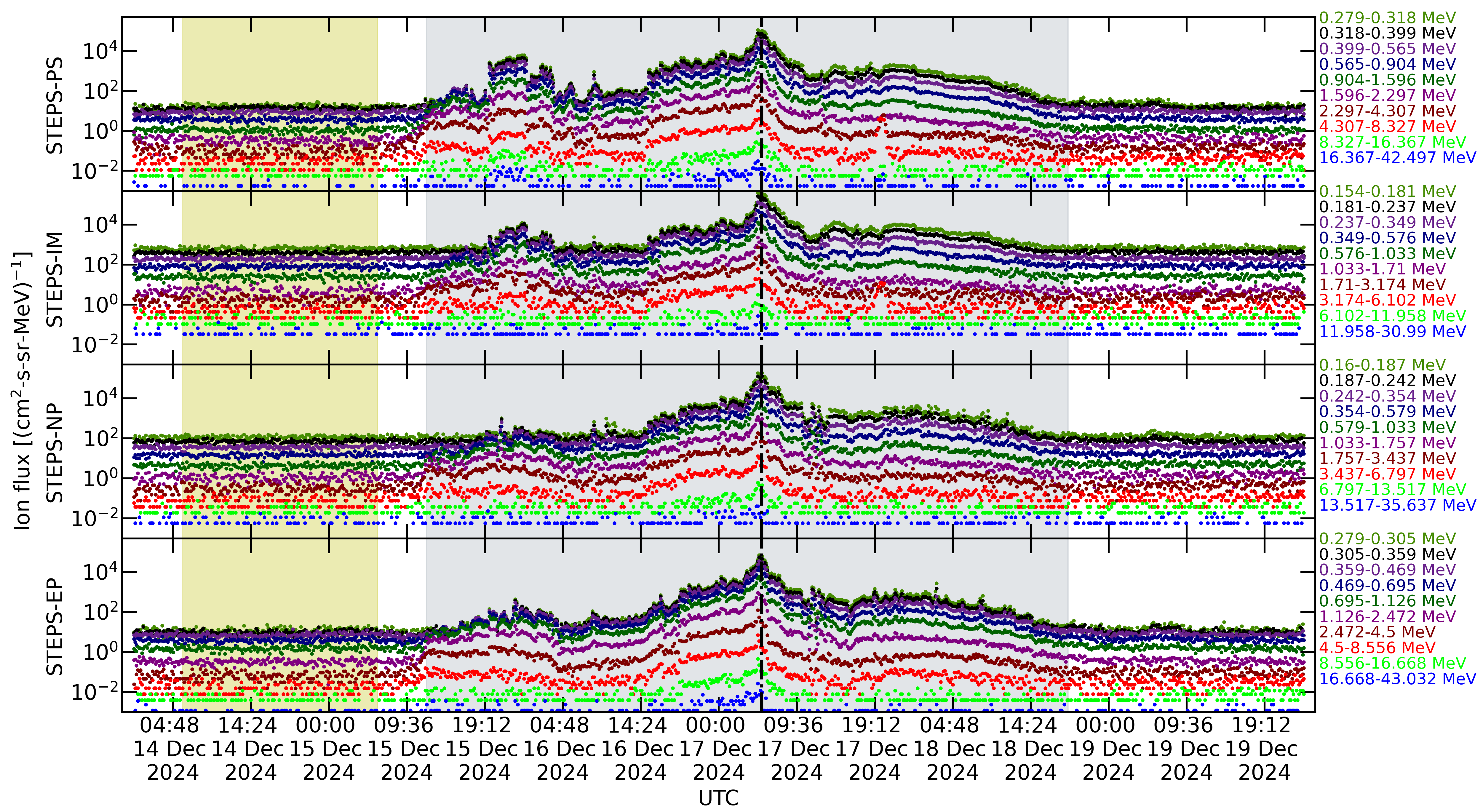}
\caption{Temporal variations of energetic ion fluxes at various energy bins as observed by (top to bottom) PS (Out), IM, NP, and Ep (Out) detectors of AL1-ASPEX-STEPS during 14 - 19 December 2024. Intervals for a quiet background and an ion enhancement event are marked by yellow and gray shaded regions, respectively. The black vertical dashed line marks the arrival time of an IP shock according to the ICME catalog by Richardson and Cane.  \label{fig_2_BD}}
\end{figure}

Figure \ref{fig_2_BD} presents the temporal variations of the ion fluxes observed by PS (outer), IM, NP, and EP (outer) detectors of AL1-ASPEX-STEPS during an interval when both quiet (yellow shaded) and event (gray shaded) time fluxes are easily identifiable. The event is due to an IP shock associated with an ICME, according to the Richardson and Cane ICME catalog (\url{https://izw1.caltech.edu/ACE/ASC/DATA/level3/icmetable2.htm#(c)}). The arrival time of the IP shock is marked by black dashed-dotted vertical line in Figure \ref{fig_2_BD}. As can be seen, the solar energetic particles (SEPs) are observed by AL1-ASPEX-STEPS sensors well before the IP shock arrives at the spacecraft location (i.e., near the L1 point). We also observe directional asymmetry in the flux variations during the event interval. This type of feature is persistently detected by AL1-ASPEX-STEPS detectors. In this work, we do not try to explore the reasons for such observations.

In addition to the temporal variations of ion fluxes, we observe differences in the spectra of ions as well. In subsequent sections, we describe some of these features.   

\begin{figure}[ht!]
\centering
\includegraphics[width=0.8\linewidth]{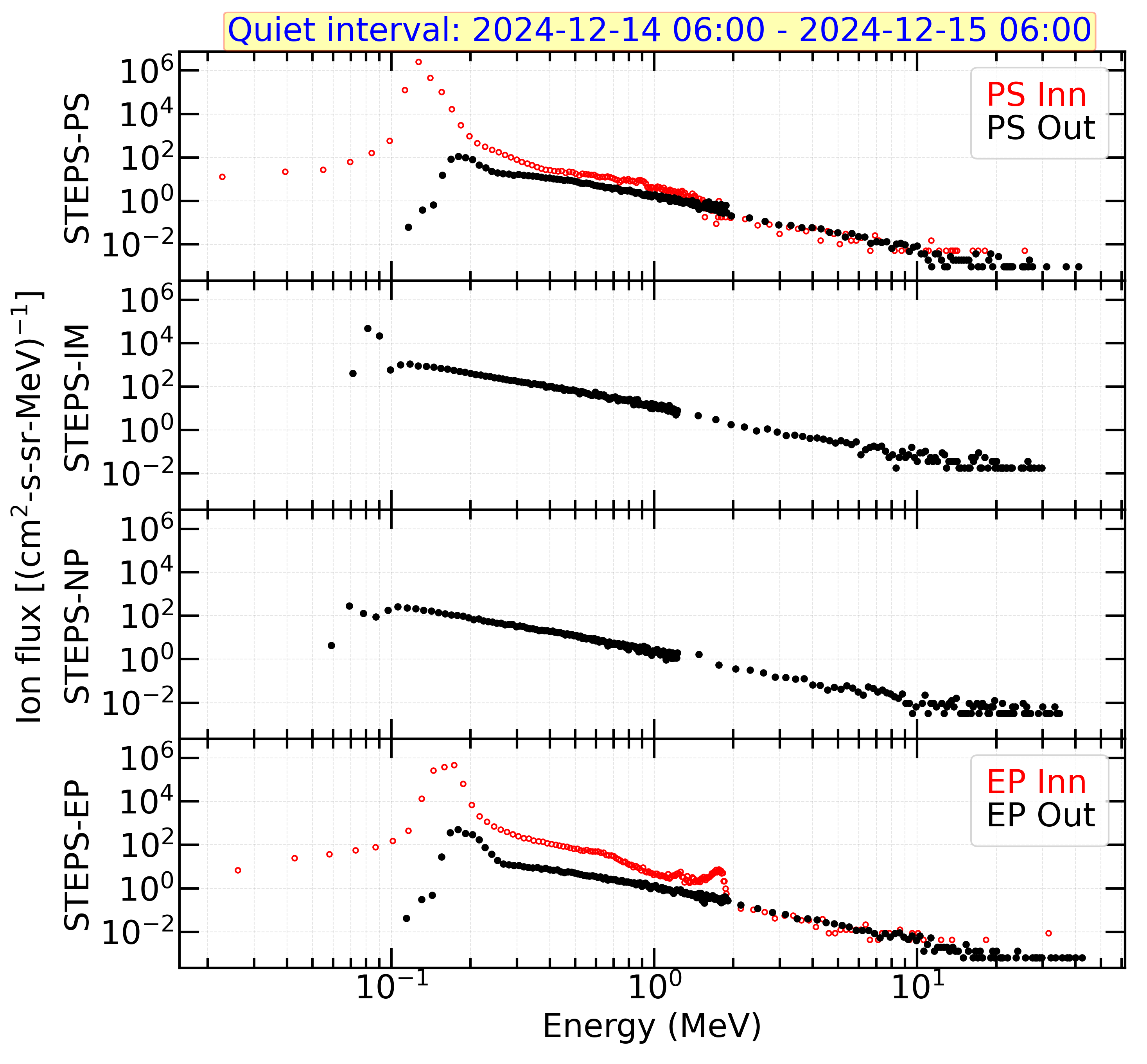}
\caption{Full energy spectra corresponding to all ADC channels of (top to bottom) PS (red: inner, black: outer), IM, NP, and EP (red: inner, black: outer) detector units of AL1-ASPEX-STEPS during the quiet time shown by yellow shaded interval in Figure \ref{fig_2_BD}. The interval is mentioned at the top of the figure. \label{fig_3_BD}}
\end{figure}

\subsection{Full energy ion spectra during quiet time}
Figure \ref{fig_3_BD} shows the full energy spectra (corresponding to all ADC channels) of the ions observed by AL1-ASPEX-STEPS sensors during the quiet (yellow-shaded) interval presented in Figure \ref{fig_2_BD}. In this figure, spectra from both the inner (Inn) and outer (Out) detectors of PS and EP units are shown in the top and bottom panels, respectively. As seen in Figure \ref{fig_3_BD}, the nature of the ion spectra for some initial ADC channels in all AL1-ASPEX-STEPS detector units differs from the rest of the spectra. This is probably due to the low-level discriminator (LLD) threshold, which is set electronically to reduce the background noise of the instrument. Below the LLD threshold, the counts in a detector is made to zero and a nonlinear behavior is observed up to a certain channel above the LLD cutoff. After the initial channels, the ion spectra observed by PS Out, IM, NP, and EP Out are linear up to $> 10 MeV$.           

Another interesting feature observed by AL1-ASPEX-STEPS is the difference in ion spectra observed by the inner and outer detectors of PS and EP units. It can be seen from the top and bottom panels of Figure \ref{fig_3_BD}, the ion spectra observed by the inner (red) and outer (black) detectors do not match below 2 MeV i.e., for the HG channels. This discrepancy is observed to be more in the case of EP units. Note that an ultra-thin dead layer has been used on the top of inner detectors (PS Inn and EP Inn), allowing comparatively low energetic ions also to fall in the detector area than in case of outer detectors. This is an interesting aspect that is being investigated currently by the payload team.        
\begin{figure}[ht!]
\centering
\includegraphics[width=0.8\linewidth]{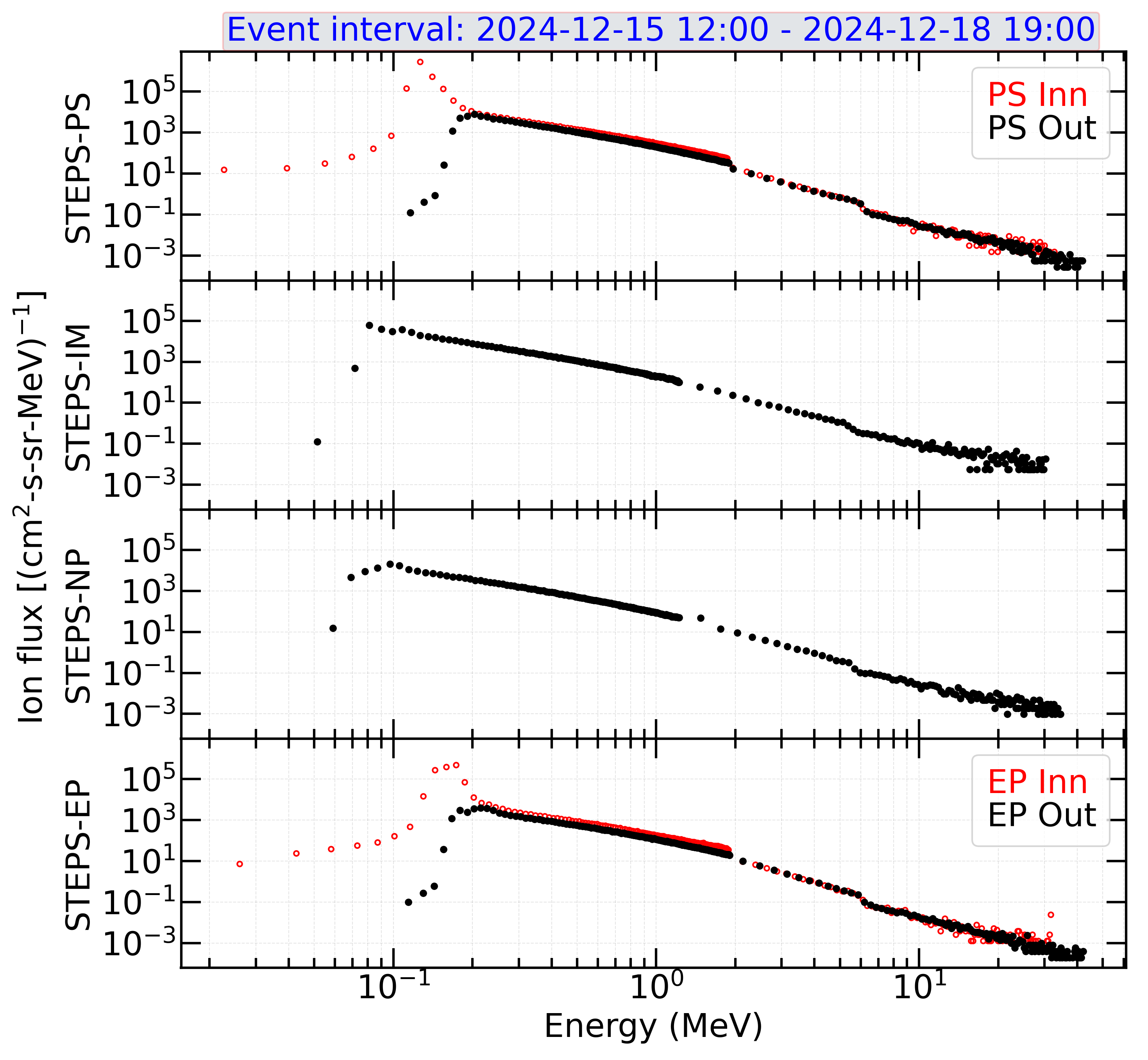}
\caption{Full energy spectra corresponding to all ADC channels of (top to bottom) PS (red: inner, black: outer), IM, NP, and EP (red: inner, black: outer) detector units of AL1-ASPEX-STEPS during the event shown by yellow shaded interval in Figure \ref{fig_2_BD}. The interval is mentioned at the top of the figure. \label{fig_4_BD}}
\end{figure}
\subsection{Full energy ion spectra during an event}
Interestingly, the inner-outer spectral discrepancy beyond the LLD cut-offs vanishes during flux enhancement events. This is clear from Figure \ref{fig_4_BD}, where we plot full energy spectra of ion fluxes averaged over the gray-shaded interval in Figure \ref{fig_2_BD}, as observed by AL1-ASPEX-STEPS sensors. As can be seen, there is a very good match between the spectra observed by inner and outer detectors of both the PS (top panel) and EP (bottom panel) units of AL1-ASPEX-STEPS throughout the entire energy ranges shown in Figure \ref{fig_4_BD}. In this work, no attempt is made to understand this aspect and we reserve this for a detailed investigation elsewhere.

A very important capability of AL1-ASPEX-STEPS is to filtering protons (H$^+$) out beyond $\approx 5.5$ MeV for single-window detectors (i.e. IM and NP) and $\approx 6.0$ MeV for dual-window detectors (i.e. PS and EP). An abrupt jump in the spectrum of each detector at a specific energy value can be observed from Figure \ref{fig_4_BD}. This is due to the fact that the effective thickness of the solid state detectors (Si-PIN) used in AL1-ASPEX-STEPS is finite. The thickness of the single-window Si-PIN detectors is 250 $\mu$m and that of the dual-window detectors is 300 $\mu$m (see \citealp{goyal2025aditya} for details). Simulation studies show that H$^+$ with energies $< \approx 6.0$ MeV is fully absorbed within the detector thickness of PS (both PS Inn and PS Out) and EP (both EP Inn and EP Out) units. This upper threshold value for IM and NP is $\approx 5.5$ MeV. Therefore, beyond the specified energies, AL1-ASPEX-STEPS detectors detect ions heavier than H$^+$. In some high-energy particle events, an additional jump is observed around 24 MeV in the PS (both PS Inn and PS Out) and EP (both EP Inn and EP Out) spectra, and around 22 MeV in the IM and NP spectra. This discontinuity is attributed to the removal of $He^{++}$ from the ion composition. Although not shown in the present analysis, this aspect will be taken up separately.

\subsection{Binned energy ion spectra during an event}
It can be seen in Figures \ref{fig_3_BD} and \ref{fig_4_BD} that ion fluxes are over-sampled in energy when these are measured in 256 ADC channels. Therefore, most of the energetic particle instruments (e.g., ULEIS \citealp{mason1998ultra}, Solar Isotope Spectrometer (SIS), \citealp{stone1998solar}, etc.) provide these data in various energy bins with, generally, increasing widths. In this way, the concerned datasets do not lose any scientific value. In case of AL1-ASPEX-STEPS also, ion flux data are provided at various energy bins.        

\begin{figure}[ht!]
\centering
\includegraphics[width=0.8\linewidth]{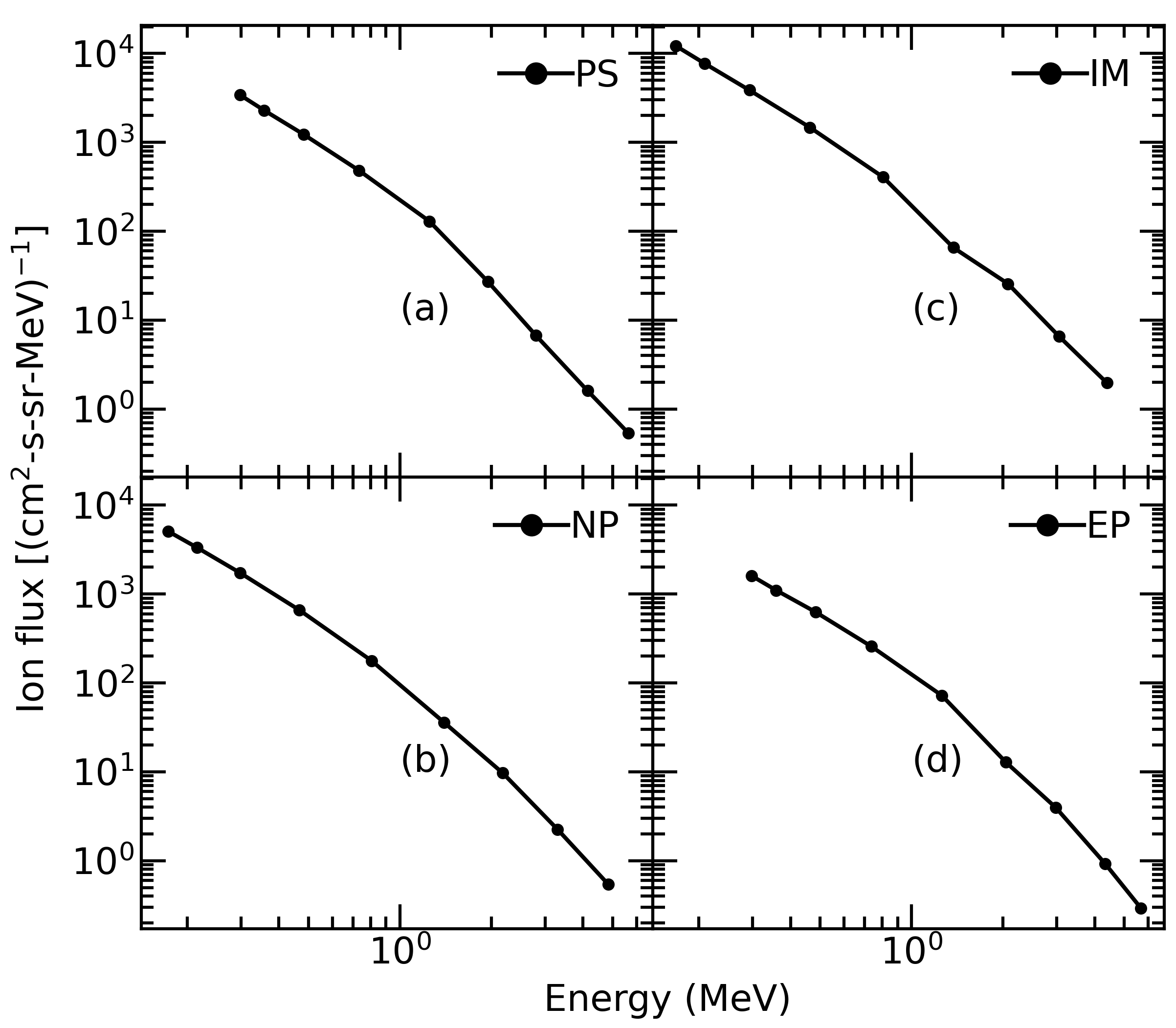}
\caption{Flux vs binned energy spectra of energetic ions observed by (a) PS (Out), (b) NP, (c) IM, and (d) EP (Out) detecors of AL1-ASPEX-STEPS during the event shown by shaded interval in Figure \ref{fig_2_BD}. \label{fig_5_BD}}
\end{figure}
Figure \ref{fig_5_BD} shows the binned energy spectra of energetic ions, as observed by PS (Out), IM, NP, and EP (Out) detectors of AL1-ASPEX-STEPS for the same event as shown in Figure \ref{fig_2_BD}. In this case, ten energy bins below $\sim$ 6 MeV (PS, EP) and $\sim$ 5.5 MeV (IM, NP) are constituted from the ADC channels and event-averaged ion fluxes are plotted corresponding to the centre points of these energy bins. It can be seen in Figure \ref{fig_5_BD} that the ion spectra do not follow single power laws up to 6 MeV during particle enhancement events like the one shown here. The transition from the suprathermal to SEP region is clearly visible in Figure \ref{fig_5_BD}. In the literature, these types of energetic ion spectra for gradual SEPs are common (see \citealp{desai2016large}) and represent different processes by which these particles are generated. Therefore, data from AL1-ASPEX-STEPS can be used for rigorous science investigations in this line.

\begin{figure}[ht!]
\centering
\includegraphics[width=0.9\linewidth]{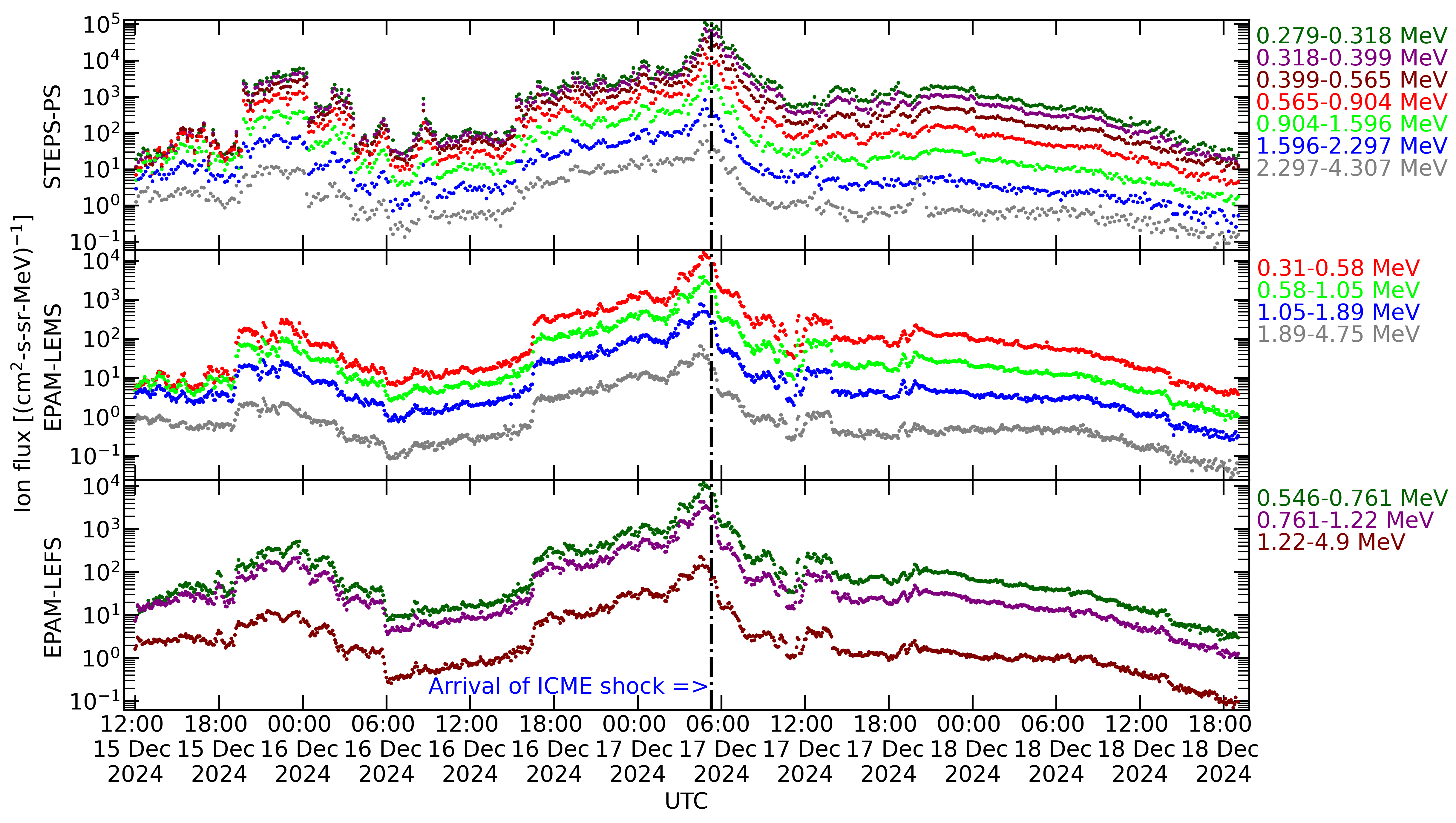}
\caption{Temporal variations of ion fluxes at different energy bands as observed by AL1-ASPEX-STEPS-PS (top), EPAM-LEMS120 (middle), and EPAM-LEFS (bottom) during 15 December 2024 12:00 UT - 18 December 2024 18:00 UT. \label{fig_6_BD}}
\end{figure}
\begin{figure}[ht!]
\centering
\includegraphics[width=0.7\linewidth]{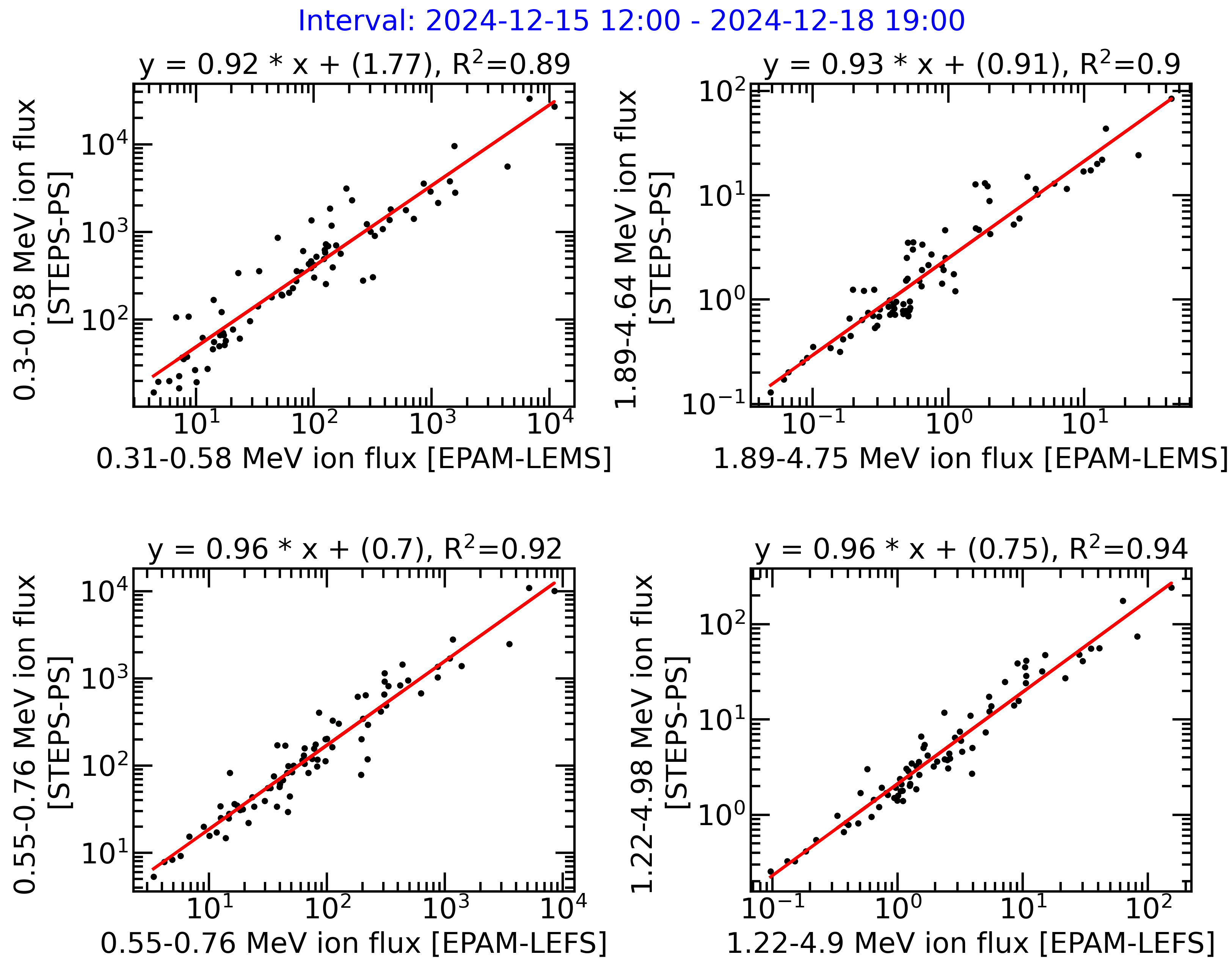}
\caption{Comparisons of hourly averaged ion fluxes observed by AL1-ASPEX-STEPS-PS (PS Out) with hourly averaged ion fluxes observed by EPAM-LEMS120 (top row) and EPAM-LEFS (bottom row) for two similar energy ranges during 15 December 2024 12:00 UTC - 18 December 2024 19:00 UTC. The red lines represent the linear fits ($y = mx + c$). The corresponding equations of the linear fits and $R^2$ values are mentioned at the top of each panel. \label{fig_7_BD}}
\end{figure}
\subsection{Comparison of AL1-ASPEX-STEPS data with ACE-EPAM data}
Validation of AL1-ASPEX-STEPS data with respect to existing datasets are provided in earlier works like \cite{goyal2025aditya}, Chakrabarty et al. 2025 (under review), and \cite{jacob2025gsics}. However, for the sake of completeness, we compare in this work the ion fluxes observed by AL1-ASPEX-STEPS-PS (outer detector) with ion fluxes observed by two detectors (Low Energy Foil Spectrometer, LEFS and Low Energy Magnetic Spectrometer, LEMS120) of EPAM \citep{gold1998electron} on-board ACE \citep{stone1998advanced} satellite at the L1 point. In this case, we choose the event interval (gray shaded) shown in Figure \ref{fig_2_BD}. Figure \ref{fig_6_BD} shows the time series plots of ion fluxes observed by AL1-ASPEX-STEPS-PS (top), EPAM-LEMS120 (middle), and EPAM-LEFS (bottom) at various energy channels. The time series plots show similar temporal variations in the ion fluxes observed by different detectors. Further, in Figure \ref{fig_7_BD}, we compare the hourly averaged ion fluxes from AL1-ASPEX-STEPS-PS with the hourly averaged ion fluxes from EPAM-LEMS120 and EPAM-LEFS telescopes for two similar energy ranges. For the comparison of AL1-ASPEX-STEPS-PS and EPAM-LEMS120, we select 0.31-0.58 MeV and 1.89 - 4.75 MeV energy bins. On the other hand, fluxes from AL1-ASPEX-STEPS-PS and EPAM-LEFS are compared for 0.55 - 0.76 MeV and 1.22 - 4.9 MeV. We calculate the Pearson's correlation coefficients ($R$) for the pairs of ion fluxes observed by AL1-ASPEX-STEPS-PS and EPAM-LEMS120 and AL1-ASPEX-STEPS-PS and EPAM-LEFS. The $R^2$ values are mentioned for each such fit (solid red lines). As can be seen, $R^2 \approx 0.9$ for all the four cases. This shows how good AL1-ASPEX-STEPS data are in comparison with data from the existing instruments like EPAM.          

\section{Conclusions} 
In summary, this article presents the one-year in-flight performance of the AL1-ASPEX-STEPS, which has been measuring suprathermal and energetic particles from the L1 point in multiple directions. Although the SR and SP detector units are currently non-operational, the instrument's general multidirectional observation capability remains intact, allowing valuable measurements in and out of the ecliptic plane. Operating in 'toggling' mode, AL1-ASPEX-STEPS has demonstrated stable and reliable performance throughout the mission period. It is capable of capturing signatures of solar energetic particles (SEPs). Current investigations focus on understanding directional asymmetries and inner-outer discrepancies in particle populations, particularly during quiet solar-wind conditions, using data from the PS and EP detector units.

The consistency of AL1-ASPEX-STEPS observations with those from other space missions further strengthens confidence in its data set. These high-quality, multi-directional measurements from the L1 vantage point offer a unique and valuable dataset for addressing key questions related to the behavior of suprathermal and energetic particles in the interplanetary medium, paving the way for future heliospheric science investigations. 

\backmatter

\bmhead{Acknowledgements}
Aditya-L1 is an observatory class mission which is fully funded and operated by the Indian Space Research Organization (ISRO). The mission was conceived and realised with the help from various ISRO centres. The science payloads and science ready data products are realised by the payload PI institutes in close collaboration with ISRO centres. 

We acknowledge the use of data from the Aditya-L1 mission of the Indian Space Research Organisation (ISRO), archived at the Indian Space Science Data Centre (ISSDC). The authors thank the ISRO science and engineering teams involved in the design, development, integration, testing, and operations of the Aditya-L1 mission, and the teams at various ISRO centres whose efforts have made this mission possible.

\bmhead{Data availability}
The AL1-ASPEX-STEPS data used in this study are publicly available through the Indian Space Science Data Centre (ISSDC) at \url{https://pradan.issdc.gov.in/al1/}. The data before the RSO phase are made available through \url{https://doi.org/10.5281/zenodo.15868882}. The ion flux data obtained from EPAM-LEMS120 and EPAM-LEFS are taken from NASA’s Coordinated Data Analysis Web (CDAWeb) at  \url{https://cdaweb.gsfc.nasa.gov/index.html}.



\bibliography{main}

\end{document}